\makeatletter
\def\ps@wanda{%
   \renewcommand{\@oddhead}{\begin{minipage}{13cm}
        \footnotesize\itshape  A slightly different version of this manuscript will appear
 in Letters in Mathematical Physics with doi \href{http://dx.doi.org/10.1007/s11005-006-0111-5}{http://dx.doi.org/10.1007/s11005-006-0111-5}.
See also \href{http://www.ma.utexas.edu/mp\_arc/c/06/06-288.pdf}{http://www.ma.utexas.edu/mp\_arc/c/06/06-288.pdf}. \end{minipage}
                                }%
   \renewcommand{\@evenhead}{}%
   \renewcommand{\@oddfoot}{\hfil\footnotesize\textsf{\bfseries\thepage}\hfil}%
   \renewcommand{\@evenfoot}{}%
}%
\makeatother
\documentclass[10pt]{amsart}
\usepackage{a4}
\usepackage{amsthm,amssymb}
\usepackage[raiselinks,colorlinks]{hyperref}
\newcommand{\be}{\begin{equation}}
\newcommand{\ee}{\end{equation}}
\newcommand{\bea}{\begin{eqnarray*}}
\newcommand{\eea}{\end{eqnarray*}}
\newcommand{\per}{{\mathop{\mathrm{per}}}}
\newcommand{\Dir}{{\mathop{\mathrm{Dir}}}}
\newcommand{\dist}{{\ensuremath{\mathrm{dist}}}}

\newcommand{\RR}{\mathbb{R}}

\newcommand{\NN}{\mathbb{N}}

\newcommand{\cD}{\mathcal{D}}
\newcommand{\cE}{\mathcal{E}}
\newcommand{\cF}{\mathcal{F}}

\newcommand{\cU}{\mathcal{U}}

\newtheorem{thm}{Theorem}

\theoremstyle{definition}

\newtheorem{clo}[thm]{Closing Remark}

\newcommand{\Hm}[1]{\leavevmode{\marginpar{\tiny%
$\hbox to 0mm{\hspace*{-0.5mm}$\leftarrow$\hss}%
\vcenter{\vrule depth 0.1mm height 0.1mm width \the\marginparwidth}%
\hbox to
0mm{\hss$\rightarrow$\hspace*{-0.5mm}}$\\\relax\raggedright #1}}}

\begin{document}
\title{Spectral gap of segments of periodic waveguides}
\author[S.~Kondej]{Sylwia Kondej$^{1,2}$}

\address[$^1$]{Institute of Physics,\, University of Zielona Gora,\, ul.~Prof.~Z.~Szafrana 4a,\, Zielona Gora,\, Poland}
\email{skondej@proton.if.uz.zgora.pl}
\author[I.~Veseli\'c]{Ivan Veseli\'c$^2$}
\address[$^2$]{Emmy-Noether-Programme of the Deutsche Forschungsgemeinschaft\vspace*{-0.3cm} }
\address{\& Fakult\"at f\"ur Mathematik,\, 09107\, TU\, Chemnitz, Germany   }

\urladdr{\href{http://www.tu-chemnitz.de/mathematik/schroedinger/quantum-wavegiudes.php}{www.tu-chemnitz.de/mathematik/schroedinger/quantum-wavegiudes.php}}

\thanks{\copyright 2006 by the authors. Faithful reproduction of this article,
         in its entirety is permitted for non-commercial purposes.%
}


\maketitle

\thispagestyle{wanda}

\begin{center}
\emph{\begin{small}Dedicated to Pavel Exner on the occasion of his 60$^{\text th}$ birthday.\end{small}}
\end{center}

\begin{abstract}
The lowest spectral gap of segments of a periodic waveguide in $\RR^2$
is proportional to the square of the inverse length.
\end{abstract}

The aim of this letter is a brief presentation of some results concerning 
spectral gaps in periodic waveguides. 
They are a representative example of the type of results derived in the
forthcoming paper \cite{KondejV-b}, see also the Closing Remark.
\medskip

Let $\gamma\colon\RR \to \RR^2$ be a $C^4$-function 
parameterised by arc-length 
and denote by $\Gamma=\gamma(\RR)$ the curve which is its range.
Assume that the curve is periodic in the following sense: there is a $p>0$
such that $\gamma(s+p)=\gamma(s)+(1,0)$ for all $s\in \RR$.
Define a periodic strip of width $\rho>0$ by $\Omega:=\{(x,y)\mid \dist\big((x,y),\Gamma\big)< \rho\}$.
Denote the normal vector $(-\dot{\gamma_2},\dot{\gamma_1})$ to $\gamma$ by $\nu$ 
and the curvature of $\gamma$ by $\kappa$.
Define the mapping  $\cF\colon \Lambda\!\!:=\RR \times (-\rho , \rho )\to 
\Omega $ by $\cF(s,u)=\gamma(s)+u\,\nu(s)$ and
assume that $\gamma$ and $\cF$ satisfy the following conditions 
\begin{equation}
\label{e-assumption}
\rho \|\kappa \|_\infty < 1 \quad \text{ and } \quad \cF \text{ is an embedding} \,.
\end{equation}
Denote by $\Lambda_L$ the segment $(-pL/2,pL/2)\times (-\rho , \rho )$ 
and by $\Omega_L$ its image $\cF(\Lambda_L)\subset \Omega$.
Let $-\Delta_\Omega$ be the Dirichlet Laplace operator in $L^2 (\Omega )$ and 
$-\Delta_{\Omega,L}$ the Laplacian in $L^2 (\Omega_L)$ with Dirichlet b.c.~on 
$\partial\Omega_L \cap \partial\Omega$ and periodic b.c.~on 
$\partial\Omega_L \setminus \partial\Omega$. 
Of course, $-\Delta_{\Omega,L}$ has purely discrete spectrum.
\medskip

The main result of this note estimates the distance between the lowest 
$E_{1,L}$ (non-degenerate) and the second $E_{2,L}$ eigenvalue of $-\Delta_{\Omega,L}$.
\begin{thm}
\label{t-gap}
There is a constant $C>0$ such that for all $L \in \NN$ satisfying $pL\ge 4\rho/\sqrt{3}$:
\[
\frac{1}{C\, L^2} \le E_{2,L}-E_{1,L} \le \frac{C}{L^2}.
\]
\end{thm}
\noindent If the curve $\Gamma$ is reflection symmetric with respect to the 
$y$-coordinate axis,
the same estimate holds if we replace the periodic part of the b.c.~by
Neumann b.c.
\smallskip

An analogous result was derived by Kirsch and Simon in \cite{KirschS-87}
for Neumann Laplacians with periodic potential, restricted to cubes. This paper was 
the motivation of the present letter. Note that due to the bound \eqref{e-comparison}
and the different behaviour of ground states near the boundary, 
Dirichlet b.c.~are harder to treat than Neumann ones.
The remainder of this letter explains the strategy of proof of 
Theorem \ref{t-gap} leaving out the technical details. 
\medskip

To analyse the waveguide Laplacian it is convenient to 
introduce a straightening transformation, see for instance \cite{KrejcirikK-05}. 
The mapping $\cF$ induces the unitary operator 
$\cU\colon L^2(\Lambda) \to L^2(\Omega)$ given by 
$\cU \phi=|G|^{-1/4}\phi \circ \cF$, where $|G|=\det G$, $G=\mathrm{diag}(h^2,1)$
and $h(s,u)=1-u\kappa (s)$.
Denote by $\cF_L$ and $\cU_L$ the restrictions of $\cF$ to $\Lambda_L$, respectively of $\cU$ 
to $L^2 (\Lambda_L)$.
Then the unitarily transformed operator $H_L:= -\cU_L^* \Delta_{\Omega,L} \cU_L$
on $L^2 (\Lambda_L)$ is given by the differential expression
\begin{equation}
\label{e-defHL}
H_L =\tilde{H}_L +V\,,\quad\text{ where }\quad
\tilde{H}_L=-\nabla G\nabla\,,\quad 
V=-\frac{\kappa^2 }{4h^2}+ \frac{\partial_s^2 h}{2h^3 }- \frac{5(\partial_s h)^2}{4h^4}\, .
\end{equation}
The domain $W_{\Dir,\per}^{2,2}(\Lambda_L)$
of $H_L$ consists of $W^{2,2}$-functions
having Dirichlet b.c.~on the part of the boundary where $u=\pm\rho$
and periodic b.c.~on the part of the boundary where $s=\pm\frac{pL}{2}$.
Of course, the spectrum of $H_L$, 
coincides with the one of $-\Delta_{\Omega,L}$.
Denote by $\cE_L$ the quadratic form corresponding to $H_L$.

The normalised eigenfunction $\psi_{1,L}$ of $H_L$ 
corresponding to $E_{1,L}$ can be chosen to be positive everywhere. 
Denote by $\Psi$ the periodic continuation of  $\psi_{1,1}$
along the $s$-coordinate axis.
It follows that $E_{1,L}= E_{1,1}$ and 
$\psi_{1,L}= (pL)^{1/2}\Psi$ for all $L\in \NN$. 
In the sequel we use the abbreviation $q:=(s,u)$.
By means of the unitary ground state transformation
\[
U\colon L^2(\Lambda_L)\to  L^2(\Lambda_L ,\psi_{1,L}^2\mathrm{d} q )\, ,
\quad \quad U f :=\psi_{1,L}^{-1} f
\]
we define the quadratic form 
\begin{equation} 
\label{e-unitaryform}
\eta _L [\phi ]:= \cE_{L}[U^{-1}\phi]- E_{1,L} \, \| U^{-1}\phi\|^2 \,,
\quad 
\phi \in \cD(\eta_{L}) =W_{\Dir,\per}^{1,2}(\Lambda_L ,\psi_{1,L}^2\mathrm{d} q)\,.
\end{equation}
Here $\|\cdot\|$ denotes the norm in $L^2(\Lambda_L)$.
The following result about the representation of a waveguide operator by a 
suitable Dirichlet form is an analog of Theorem 4.4 (and C.1) in \cite{DaviesS-84}.

\begin{thm}
\label{t-representation} 
The quadratic form $\eta_L$ admits the following representation
\begin{equation} 
\label{e-formrep}
\eta_L[\phi] =
\int_{\Lambda _L} (G\nabla\phi)\cdot (\nabla \overline{\phi}) \psi_{1,L}^2 \mathrm{d} q\,, 
\quad \mathrm{for }\quad \phi \in \cD(\eta_L)\,.
\end{equation}
\end{thm}

From the above theorem we directly obtain
\begin{multline}\label{e-gap}
E_{2,L}- E_{1,L}=
\\=\inf \Big\{ \eta_L[\phi]
\,\Big|\,
\phi \in \cD(\eta_L)\, , \int_{\Lambda _L} |\phi|^2  \psi_{1,L}^2 \mathrm{d} q=1\,
\,, \, \int_{\Lambda _L} \phi  \psi_{1,L}^2 \mathrm{d} q=0 \Big\}\,.
\end{multline}
Following the reasoning of \cite{KirschS-87},  equality 
\eqref{e-formrep} allows us to bound the gap
$E_{2,L}- E_{1,L}$ in terms of the first two eigenvalues 
$\tilde{E}_{2,L}\,,\tilde{E}_{1,L}$ of the comparison operator 
$\tilde{H}_L$ defined in \eqref{e-defHL} and its ground state 
$\tilde{\psi}_{1,L}$.
For  $L\in \NN$ set
$a_+^L =\max_q \frac{\tilde{\psi}_{1,L}(q)}{\psi_{1,L}(q)}\,,
a_-^L =\min_q \frac{\tilde{\psi}_{1,L}(q)}{\psi_{1,L}(q)}$
and note that by periodicity we have $a_\pm^L=a_\pm^1=:a_\pm$.

\begin{thm}
\label{t-comaprison} For all $L\in \NN$ we have
\begin{equation}
\label{e-comparison}
\left( \frac{a_-}{a_+} \right)^2 (\tilde{E}_{2,L}- \tilde{E}_{1,L})
\leq  E_{2,L}- E_{1,L} \leq \left( \frac{a_+}{a_-} \right)^2 
(\tilde{E}_{2,L}- \tilde{E}_{1,L})\,.
\end{equation}
\end{thm}

To apply the theorem, it is necessary to know how the ground states of the two operators
behave near the boundary of $\Lambda_1$.
Using Theorem 9.2 from \cite{DaviesS-84} one can show that $a_+$ and $a_-$ 
are finite and positive. 
The quoted Theorem applies only to Dirichlet b.c.~on smooth domains in $\RR^d$,
whereas $\Lambda_1$ has a boundary with corners and $H_1, \tilde H_1$ have boundary segments 
equipped with periodic b.c.
This problem is eliminated by mapping $\Lambda_1$ to an annulus in $\RR^2$.
\medskip

It remains to estimate the distance $\tilde E_{2,L}-\tilde E_{1,L}$.
To this aim we compare it with the first spectral gap 
of the Laplacian $-\Delta_{\Lambda,L}$ in $L^2 (\Lambda_L)$.
Due to the assumption $pL\ge 4\rho/\sqrt{3}$, this gap equals $4\pi^2 \, (pL)^{-2} $. 
The operators $\tilde{H}_L$ and $-\Delta_{\Lambda,L}$ have the same ground state, namely
$\tilde{\psi}_{1,L}(s,u)=\big(\frac{\pi}{2\,\rho\, p\,L}\big)^{1/2} \cos \frac{\pi u}{2\rho}$.

Formula \eqref{e-gap} holds for $\tilde H_L$ if we replace 
$\psi_{1,L}$ by $\tilde\psi_{1,L}$ and for $-\Delta_{\Lambda,L}$ 
if we replace  $\psi_{1,L}$ by $\tilde\psi_{1,L}$ and 
$G$ by $\bigl( \begin{smallmatrix}1&0\\ 0&1\end{smallmatrix} \bigr)$.
To obtain uniform pointwise bounds on $G$, note that $h$ has uniform positive upper and lower bounds,
due to assumption \eqref{e-assumption}.
Thus there exists $c>0$ such that 
\[
c^{-1} L^{-2}\leq \tilde E_{2,L}-\tilde E_{1,L}\leq c L^{-2}\,.
\]
Combining the above inequality with \eqref{e-comparison} 
we complete the proof of Theorem~\ref{t-gap}.
\smallskip

\begin{clo}
In the forthcoming paper \cite{KondejV-b} we give all details of the proofs 
in this note, 
discuss the relation to other results obtained in the literature, and
address the extension to the following more general periodic operators:
\begin{list}{-}{\itemindent -1em}
\item waveguides in three or more dimensions,
\item layers in three or more dimensions,
\item presence of a (periodic) potential in the original operator,
\item waveguides and layers with Neumann b.c., and
\item waveguides and layers with relaxed regularity 
and symmetry conditions.
\end{list}
Furthermore, we prove general, abstract analogs of Theorems~\ref{t-representation} 
and \ref{t-comaprison}
for divergence form operators on waveguides. These results are applied to 
derive estimates on the asymptotics of the 
density of states at the minimum of the spectrum of a periodic waveguide or layer.
\smallskip

We close this letter by noting that related estimates on eigenvalues of 
waveguides were obtained in \cite{Yoshitomi-98} and \cite{BorisovE-04,Borisov-06}. 
In \cite{Yoshitomi-98} the Floquet-Bloch spectrum of thin periodic waveguides was analysed,
whereas \cite{BorisovE-04,Borisov-06} consider a pair of straight waveguides with coupling
through windows in the common boundary. 
\end{clo}

It is a pleasure to thank D.~Krej{\v{c}}i{\v{r}}{\'{\i}}k for helpful comments
and H.~Vogt for enlightening discussions. This research was supported by the DFG 
under grants Ve 253/2-1, 2-2 and 3-1.

\thispagestyle{empty}

\def\cprime{$'$} \def\cprime{$'$}

\end{document}